\documentclass[
    ,final            
  ]
  {aipproc}

\layoutstyle{6x9}

\newcommand{\J}{J/\psi}

\newcommand{\ccbar}{c \bar c}
\newcommand{\Nccbar}{N_{\ccbar}}

\newcommand{\pts}{\langle{p_T}^2\rangle}

\newcommand{\pt}{$p_T$}

\begin{document}

\title{In-Medium formation of J/Psi as a probe of charm quark thermalization
                                                         }


\author{R. L. Thews}{
  address={Department of Physics, University of Arizona, Tucson, AZ 85721 USA}
}

\begin{abstract}
Charmonium formation via charm quark in-medium recombination in  
heavy ion interactions at collider energies has 
the potential to probe some properties of the medium by utilizing the 
sensitivity of the recombination process to the
momentum distribution of the quarks.  We have examined the 
 transverse momentum spectra of $\J$, characterized by $\pts$, which result
from the formation process in which the 
charm quark distributions are unchanged from their initial production
in a pQCD process. This is contrasted with the case in which  the charm
quarks have completely come into thermal equilibrium with an expanding medium
whose properties are determined by the spectra of produced light hadrons.   
We find that the resulting $\pts$ of the formed $\J$ provide a distinct signature 
of the underlying charm quark spectra, and that signature is essentially
independent of the detailed dynamics of the in-medium formation
reaction.  In addition, both of these signatures are sufficiently 
separated from the case in which no in-medium formation takes place.
Finally, utilizing a model for the fraction of $\J$ which originate
from in-medium formation, we predict the centrality behavior of
these signatures.
\end{abstract}

\maketitle


\section{}
The role of $\J$ produced in high energy heavy ion collisions as a signature
for color deconfinement \cite{Matsui:1986dk} has evolved in recent years with the 
realization that at collider energies an additional formation mechanism may
become significant \cite{Braun-Munzinger:2000px,Thews:2000rj}.
This depends on the initial production of multiple $\ccbar$ pairs 
in sufficient numbers.  Initial estimates \cite{Gavai:1994in}
from extrapolation of fixed-target
data put this number at about 10 for central Au-Au collisions at RHIC.  
Subsequently, measurements by the PHENIX and STAR experiments indicate 
even higher numbers, between 20 \cite{Adler:2004ta} and 40 \cite{Adams:2004fc},
respectively.  The in-medium formation picture we consider here 
uses competing formation
and dissociation reactions in a Boltzmann equation to calculate the final $\J$ 
population.  The absolute value of this formation was found to be 
very sensitive to the underlying
charm quark momentum distributions \cite{Thews:2001hy}.  In addition, there
is significant dependence on 
details of the size and expansion profile of the deconfinement
region, for which various model parameters must be introduced.  
The initial PHENIX data \cite{Adler:2003rc}
suffered from low statistics, and was compatible with 
a fairly large region of model parameter space \cite{Thews:2003da}.

Recent work in this area concentrated on finding a signature 
for in-medium $\J$ formation which is
independent of the detailed dynamics and magnitude of
the formation.  We found that the \pt spectrum of the 
formed $\J$ may provide such a signature \cite{Thews:2005vj}.
Our first calculations 
of in-medium formation used initial charm quark 
momentum distributions from NLO pQCD amplitudes to generate a sample
of $\ccbar$ pairs, supplemented
by an initial-state transverse momentum kick to simulate confinement and 
nuclear effects.  
In the absence of in-medium formation, \pt of the $\J$ follows from that
of the initially-produced individual $\ccbar$ pairs, but one must use
a model for the magnitude of the hadronization process. For the 
\emph{normalized} \pt spectrum, we start with that of the pair spectrum.
In the evolution of the interacting
system size from pp to pA to AA collisions, the \pt will be
in general increased due to initial-state effects of interaction
of constituents in the nuclei. One can express this effect as

\begin{equation}
\pts_{pA} - \pts_{pp}\; = \lambda^2\; [\bar{n}_A - 1], 
\label{pApt}
\end{equation}
where $\bar{n}_A$ is the impact-averaged number of inelastic interactions
of the projectile in nucleus A, and $\lambda^2$ is the square of the
transverse momentum transfer per collision. For a nucleus-nucleus collision,
the corresponding relation is
\begin{equation}
\pts_{AB} - \pts_{pp}\; = \lambda^2 \;[\bar{n}_A + \bar{n}_B- 2]. 
\label{AApt}
\end{equation}
The PHENIX measurements of $\J$ \pt spectra in pp and minimum-bias 
d-Au interactions \cite{Adler:2005ph,deCassagnac:2004kb}
allow us to determine the amount of initial state $k_T$ needed to
supplement our collinear pQCD events.  One can then extrapolate to
Au-Au and predict the spectrum of $\J$ which are produced from 
hadronization of the initial "diagonal" $\ccbar$ pairs,
again for minimum bias interactions.  (We use diagonal
to distinguish these pairs from the "off-diagonal" combinations which
contribute to in-medium $\J$ formation.)
One finds
$\bar{n}_A = 5.4$ for minimum bias d-Au interactions at RHIC
energy (using $\sigma_{pp}$ = 42 mb), which leads to
$\lambda^2 = 0.35 \pm 0.14\ GeV^2$.  We note that the relatively large
uncertainty comes entirely from the difference in $p_T$ broadening
in d-Au between positive and negative rapidity.

\begin{figure}[h]
  \includegraphics[clip, height=.45\textheight]{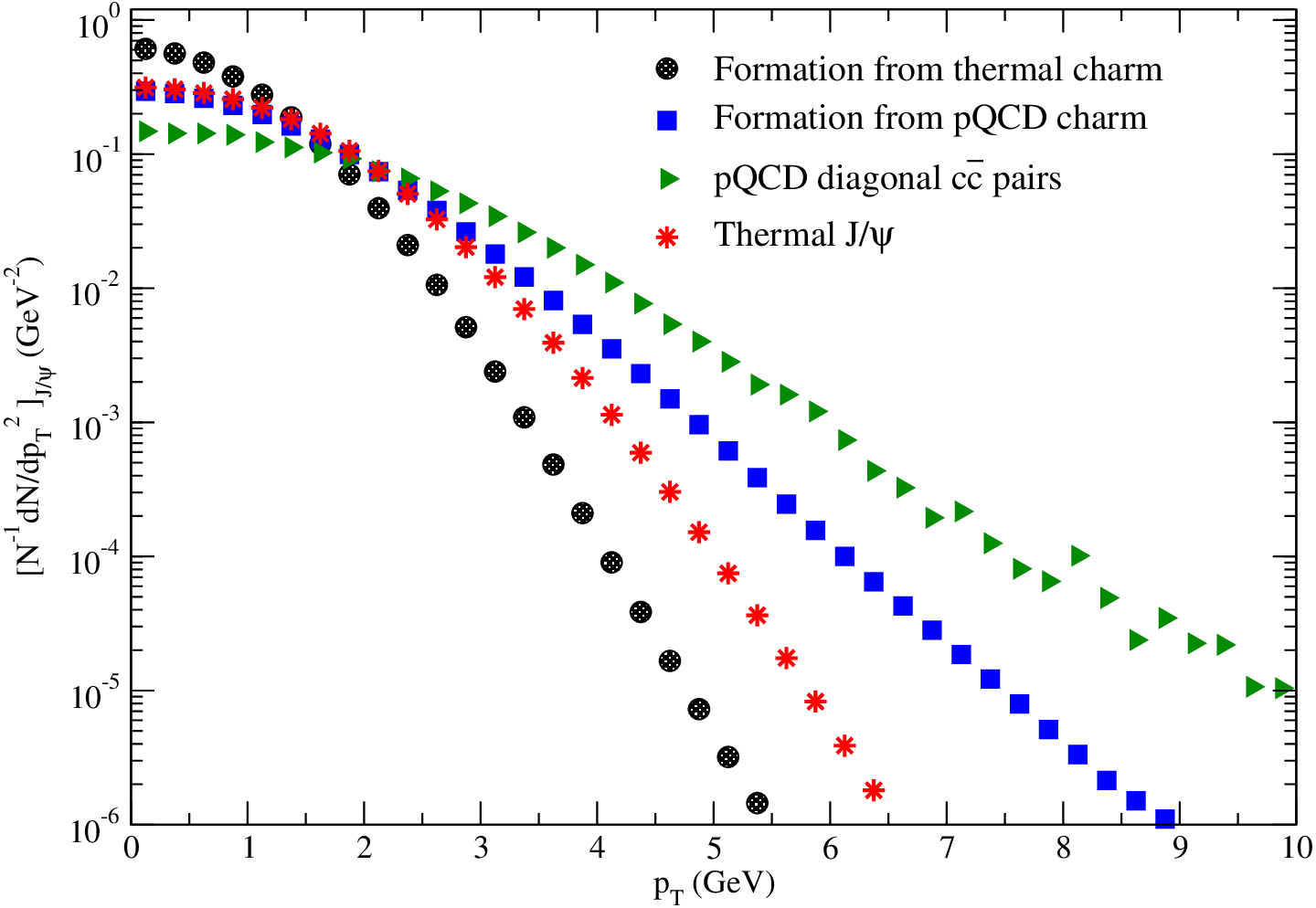}
  \caption{Comparison of im-medium $\J$ transverse momentum spectra
           predictions from various scenarios.}
\label{jpsiptallpredictions}
\end{figure}

Our prediction for the "normal" evolution of the \pt spectrum in Au-Au
interactions is shown by the triangular points in Fig.\ref{jpsiptallpredictions}.
There is
some scatter at high \pt due to the limited statistics of the number
of pQCD-generated pairs.  The $\pts$ of this spectrum is approximately
6.3 $GeV^2$.

The properties of the normalized pt spectrum for the formation process  
follow from two separate effects:  First, the fact that the process is dominated by
the off-diagonal pairs introduces a modified initial \pt distribution.
Next, one weights these pairs by a formation probability for $\J$.  We use
the operator-product motivated cross section \cite{Peskin:1979va,Bhanot:1979vb} 
for $\ccbar$ forming $\J$
with emission of a final-state gluon, which of course is just the
inverse of the dissociation process.  However, any cross section which
has the same general properties as this one gives essentially the
same result \cite{Thews:2005vj}.
We show by the square points in 
Fig.\ref{jpsiptallpredictions} the prediction for the formed $\J$.
 One sees that this spectrum, characterized by $\pts$ approximately 3.6 $GeV^2$,
 is substantially narrower
than the one with no in-medium formation. 

We next considered charm quark momentum distributions which would
follow if the charm quark interaction with the medium were so 
strong that they come into thermal equilibrium with the expanding
region of deconfinement.  The parameters of temperature and 
maximum transverse expansion rapidity are determined by a fit to
this thermal behavior of the produced light hadrons. The application 
to charm quarks was originally motivated in Ref. \cite{Batsouli:2002qf}, 
who showed that
the low-\pt spectrum of decay leptons from charmed hadron decays would not
be able to differentiate between the thermal and a purely pQCD
distribution.   We show here, however, that the \pt spectrum of
in-medium formed $\J$ is very sensitive to this distribution \cite{Greco:2003vf}.  
The
circles in Fig.\ref{jpsiptallpredictions} result from formation 
calculations using T = 170 MeV and
$y_{Tmax}$ = 0.5 for the thermal charm quarks.
One sees that this \pt spectrum
is narrower yet than in-medium formation from pQCD quarks, with
$\pts$ approximately 1.3 $GeV^2$.  Finally, we show by the stars 
the \pt spectrum of $\J$ which themselves obey this thermal
distribution.  The resulting spectrum falls between the in-medium 
formation spectra for either pQCD or thermal charm quark momentum 
distributions, with $\pts$ approximately 3.0 $GeV^2$.  

\begin{figure}[h]
  \includegraphics[clip, height=.45\textheight]{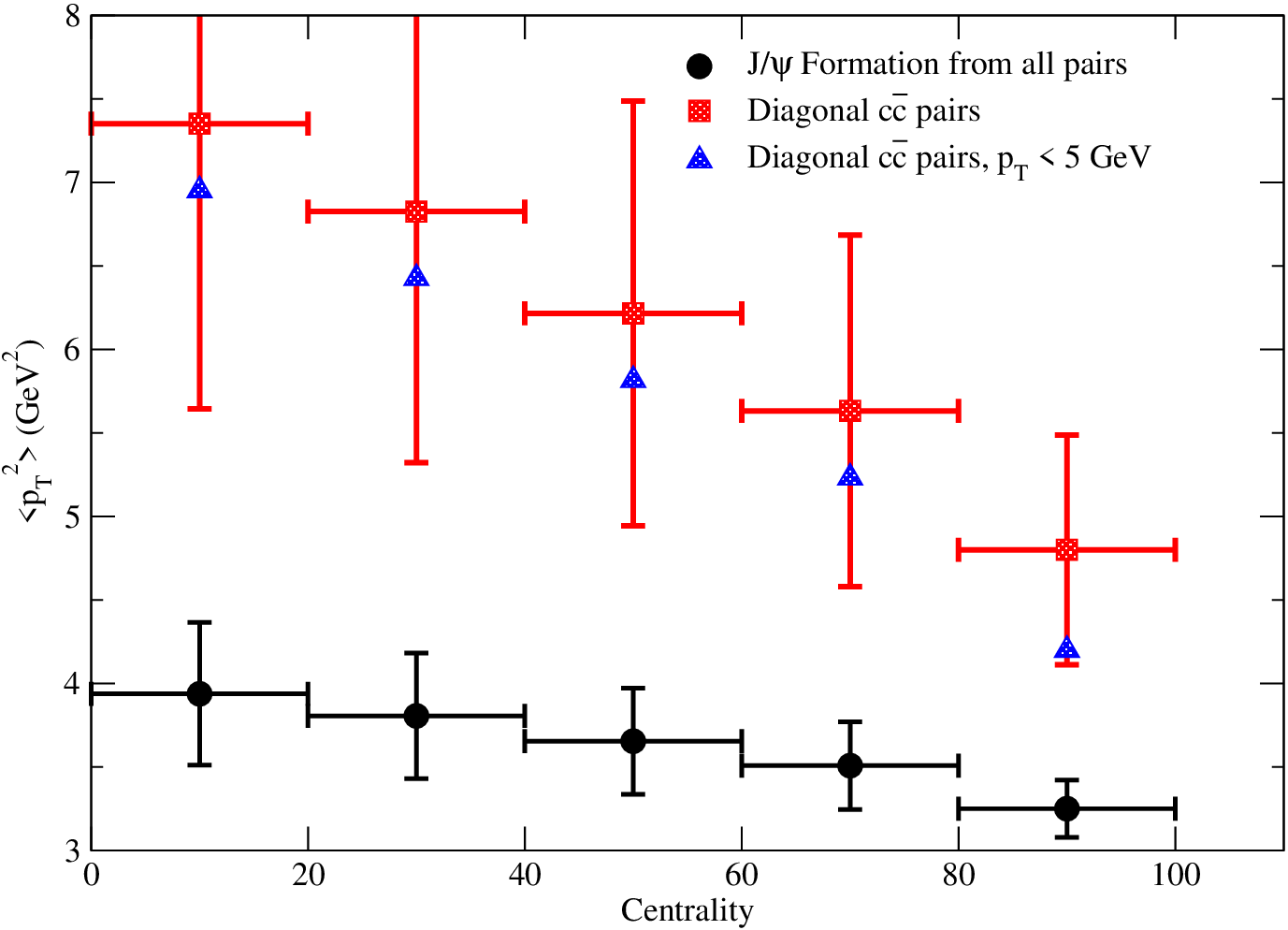}
  \caption{Centrality dependence of $\J \pts$ contrasting predictions 
           assuming either 100 \% production from initial $\ccbar$ 
           pairs or 100 \% in-medium formation.}
\label{jpsiptwidthsvscentrality}
\end{figure}

We now proceed to investigate the variation of the pQCD-based results with
respect to the collision centrality in Au-Au interactions.  First, we
use the value of $\lambda^2$ extracted from pp and pA data, together with
values of $\bar{n}_A$ calculated as a function of collision centrality, to 
recalculate the $\pts$ values for either the initial production or the
in-medium formation separately.  This provides the centrality behavior of
the $\J$ spectrum in the case that one or the other of these mechanisms is
solely responsible for the total $\J$ population.  We show these results together 
in Fig. \ref{jpsiptwidthsvscentrality}. 
One sees as expected that $\pts$ is maximum for the most central 
collisions, but the absolute magnitudes are widely separated for
initial production and in-medium formation at each centrality.  One should note
that the uncertainties are dominated by the difference between the 
\pt-broadening measurements at positive and negative rapidities
in the d-Au interactions.  Thus the point-to-point uncertainties
are much smaller for the centrality behavior.  We have also
included separate values for $\pts$ in the region limited by
\pt < 5 GeV, to facilitate comparison with experiment in this same range.

\begin{figure}[h]
  \includegraphics[clip, height=.45\textheight]{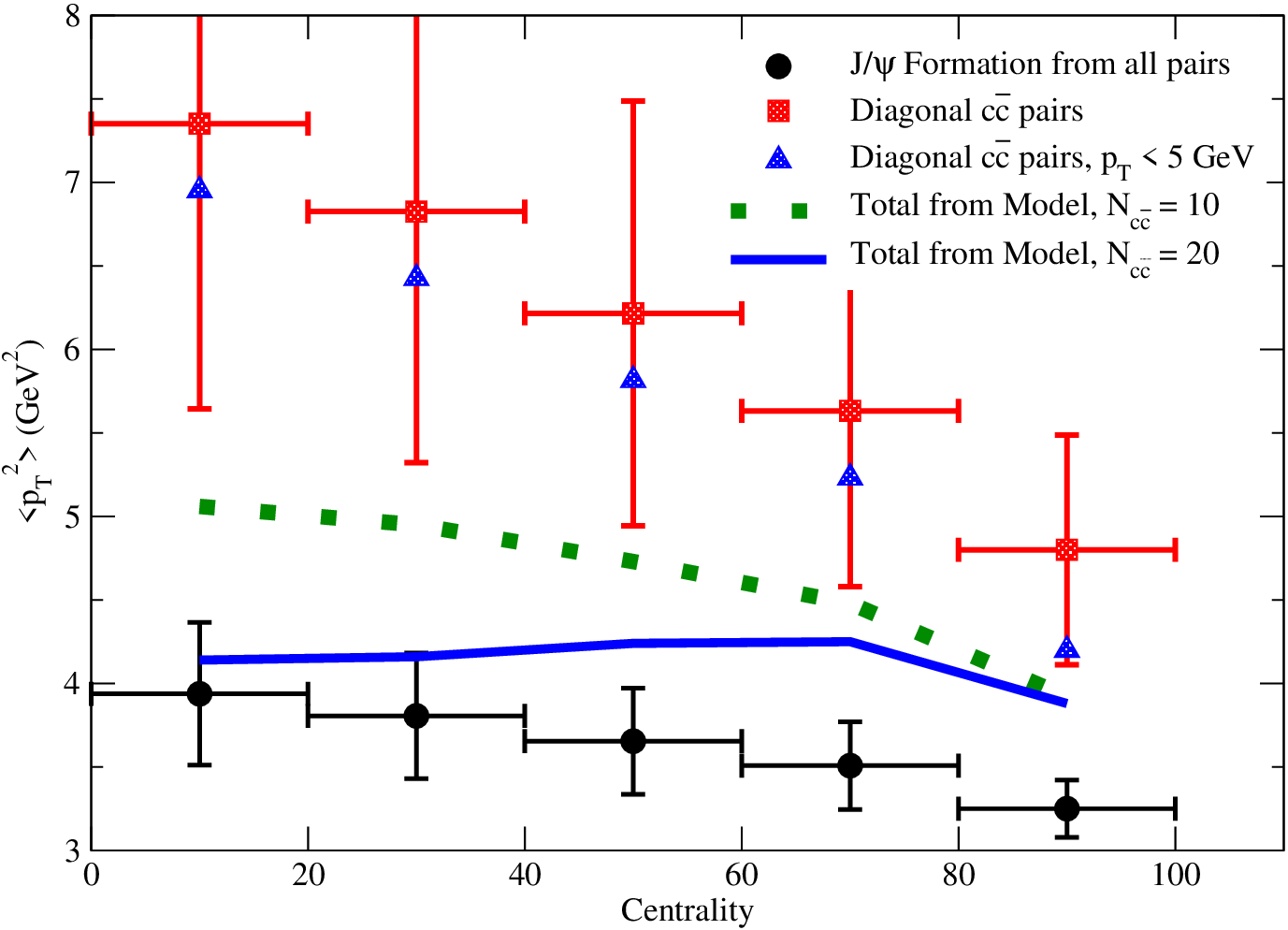}
  \caption{Centrality dependence of $\J \pts$ using a model calculation
           to estimate the relative contributions of initial production 
           and in-medium formation.} 
\label{jpsitotalptwidthvscentrality}
\end{figure}

In order to provide a meaningful prediction for the overall $\J$
spectrum, one should of course include both initial production and
in-medium formation together as sources.  This requires some estimate of the
relative magnitudes of these processes, and is subject to considerable
model uncertainties.  What we can say, however, is that in-medium formation
will be most dominant for central collisions, where the quadratic dependence
on $\Nccbar$ is enhanced.  Conversely, one expects that initial production
will increase in relative importance for very peripheral collisions.  To get
an approximate idea of how this effect will appear, we revert to our
original model calculations which included the absolute magnitude results 
\cite{Thews:2001hy}.  The relevant parameter is the number of initial
pairs, $\Nccbar$, parameterized by its value at zero impact parameter.
These results are
shown in Fig. \ref{jpsitotalptwidthvscentrality} for two representative
values of $\Nccbar$ = 10 and 20.
The anticipated dominance of each at opposite ends of the centrality scale
is seen to be realized.  In addition,  since the centrality dependence 
of each separate contribution is in the same direction, the dependence of
the total $\pts$ is somewhat more flat than either of them separately.

\begin{figure}[h]
  \includegraphics[clip, height=.45\textheight]{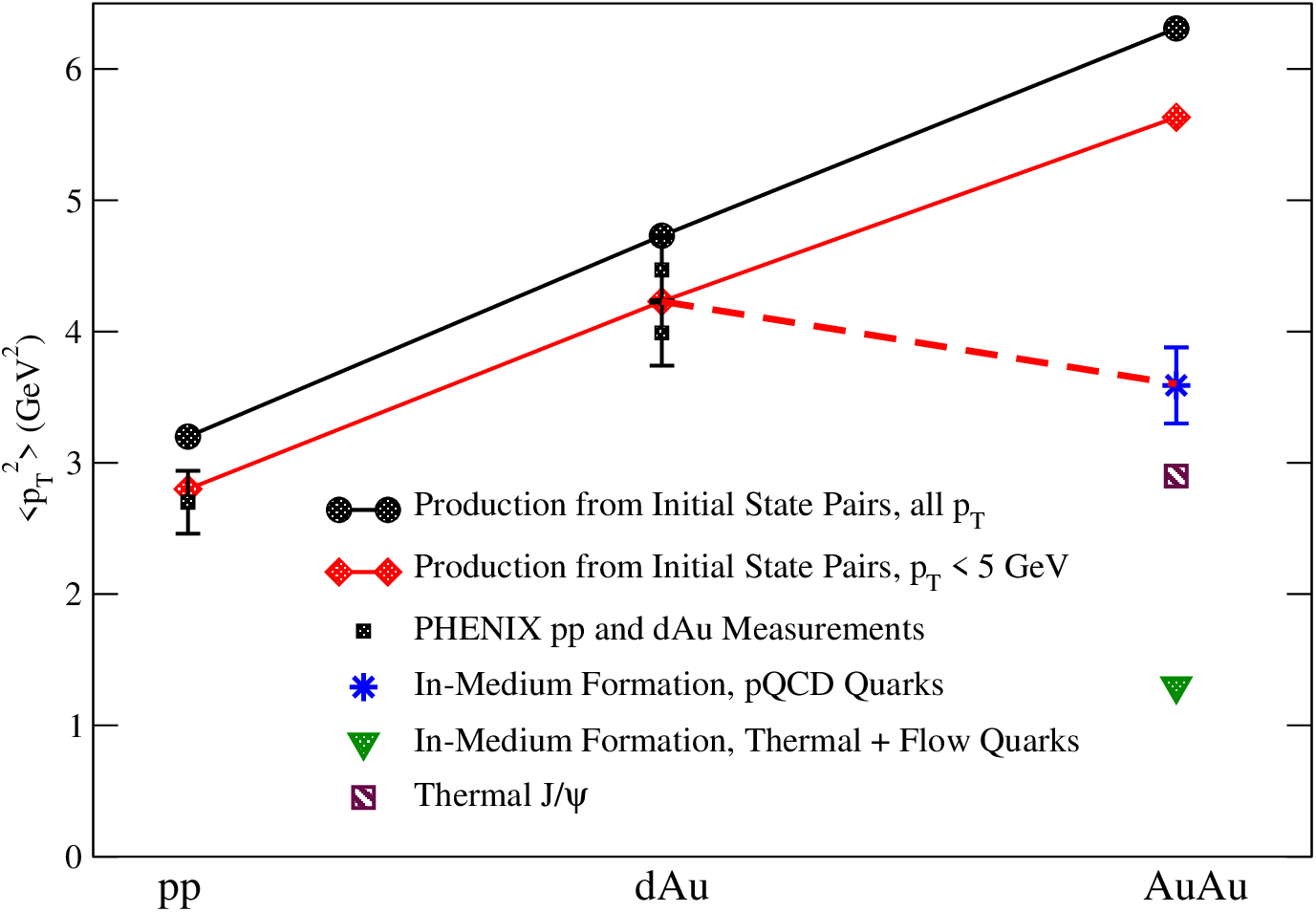}
  \caption{Comparison of initially-produced $\J \pts$ with in-medium
           formation predictions for minimum bias collisions.}
\label{ptwidthevolution}
\end{figure}

Finally, we include a graphical representation of the evolution of  
$\pts$ as a function of system size in Figure \ref{ptwidthevolution}.  
For simplicity of presentation, we revert back to the minimum bias case. 
In the absence of in-medium formation, the behavior of $\pts$ is monatonic
increasing with system size.  Including in-medium formation as a significant
component for minimum bias collisions leads to a reduced $\pts$ in
Au-Au interactions.  The numerical results indicate that this reduction
is such that the predicted Au-Au value is even below
that measured in d-Au, i.e. a non-monatonic behavior.  Hence we claim that
observation of such a non-monatonic behavior can be taken as a signal of
in-medium formation, while the absolute value of such a behavior will be
correlated with the specific charm quark distribution in the medium.


\begin{theacknowledgments}
This research was partially supported by the U.S. Department of
Energy under Grant No. DE-FG02-04ER41318.
\end{theacknowledgments}


\end{document}